\documentclass[preprint,tightenlines,preprintnumbers,amsmath,amssymb,nofootinbib]{revtex4}

\usepackage{epsfig,latexsym,cancel,amssymb,amsmath,mathrsfs,verbatim}
\usepackage{color}
\usepackage{graphicx}

\newcommand{\be}{\begin{equation}}
\newcommand{\ee}{\end{equation}}
\newcommand{\bea}{\begin{eqnarray}}
\newcommand{\eea}{\end{eqnarray}}
\newcommand{\ba}{\begin{array}}
\newcommand{\ea}{\end{array}}

\long\def\symbolfootnote[#1]#2{\begingroup%
\def\thefootnote{\fnsymbol{footnote}}\footnote[#1]{#2}\endgroup} 

\begin{document}

\preprint{MCTP-11-37}

\title{$D$ vs $d$: CP Violation in Beta Decay \\and Electric Dipole Moments}

\author{John Ng} 
\affiliation{TRIUMF Theory Group, 4004 Wesbrook Mall, Vancouver, BC, Canada, V6T 2A3}
\author{Sean Tulin}
\affiliation{Michigan Center for Theoretical Physics, Department of Physics,
University of Michigan, Ann Arbor, MI, USA, 48109}

\date{\today}% It is always \today, today,

\begin{abstract}

The T-odd correlation coefficient $D$ in nuclear $\beta$-decay probes CP violation in many theories 
beyond the Standard Model.  We provide an analysis for how large $D$ can be in light of constraints from electric dipole moment (EDM) searches.
We argue that the neutron EDM $d_n$ currently provides the strongest constraint on $D$, which is $10 - 10^3$ times stronger than current direct limits on $D$ (depending on the model).  In particular, contributions to $D$ in leptoquark models (previously regarded as ``EDM safe'') are more constrained than previously thought.
Bounds on $D$ can be weakened only by fine-tuned cancellations or if theoretical uncertainties are larger than estimated in $d_n$.  We also study implications for $D$ from mercury and deuteron EDMs.

\end{abstract}

%\pacs{}
\maketitle

\section{Introduction}

The search for CP violation beyond the Standard Model (SM) remains an open question at the forefront of nuclear physics, particle physics, and cosmology.\footnote{The discrete symmetries discussed herein are charge conjugation (C), parity (P), and time reversal (T) symmetries.  Assuming CPT invariance, T violation is equated with CP violation.}
New CP violation is a generic feature of physics beyond the SM~\cite{Grossman:1997pa}, and is likely required to explain the baryon asymmetry of the Universe~\cite{Riotto:1999yt}.  
Furthermore, unlike the SM Kobayashi Maskawa (KM) phase~\cite{Kobayashi:1973fv}, new CP violation may be unconnected with flavor and can be probed in systems of ``ordinary matter'' through searches for T violation in nuclear $\beta$-decay and electric dipole moments (EDMs) of atoms, nucleons, and nuclei.

CP violation in $\beta$-decay is manifested through T-odd triple product correlations~\cite{Jackson:1957zz}.  (See Refs.~\cite{Deutsch:1996qm,Herczeg:2001vk,Severijns:2006dr,Nico:2006yg} for reviews of fundamental symmetry tests in $\beta$-decay.)  In this work, we study the so-called $D$ correlation, corresponding to the triple product \mbox{$\langle \mathbf{J} \rangle \cdot \mathbf{p}_e \!\times\! \mathbf{p}_\nu$}, where $\langle\mathbf J \rangle$ is nuclear polarization, and $\mathbf{p}_{e}$ ($\mathbf{p}_{\nu}$) is the $e^\pm$ ($\nu$) momentum.  It is useful to write $D \equiv D_t + D_f$ to delineate fundamental T violation ($D_t$) from T-even final state 
effects ($D_f$)~\cite{Herczeg:2001vk}.  In the SM, the KM phase contribution to $D_t$ is vanishingly small~\cite{Herczeg:1997se}.  Therefore, to the extent that $D_f$ is computable or negligible, measurements of $D$ directly probe CP violation beyond the SM.

To date, $D$ has been measured for the neutron~\cite{Steinberg:1974zz,Erozolimsky:1974zz,Erozolimsky:1978zi,Lising:2000pa,Soldner:2004xm,Mumm:2011nd} and $^{19}\textrm{Ne}$~\cite{Baltrusaitis:1977em,Hallin:1984mr}.  The best neutron $D$ measurement has been obtained recently by the emiT collaboration~\cite{Mumm:2011nd}:\footnote{We have added in quadrature statistical and systematic errors quoted in Ref.~\cite{Mumm:2011nd}.}
\be
D_n = (-1.0 \pm 2.1) \times 10^{-4}  \; .
\ee
Final state interactions give $D_{f}= \mathcal{O}(10^{-5})$~\cite{Callan:1967zz}, and have been computed to an accuracy better than 1\%~\cite{Ando:2009jk}.  Although $D_n$ measurements so far agree with SM expectations, there remains (in principle)
a discovery window for future experiments down to $D_n \sim 10^{-7}$.  For $^{19}\textrm{Ne}$, an average of previous measurements~\cite{Baltrusaitis:1977em,Hallin:1984mr} gives
\be
D_{\textrm{Ne}} = (1\pm6)\times 10^{-4} \; ,
\ee
which has reached a level comparable to final state interaction effects $D_f \sim 10^{-4}$~\cite{Hallin:1984mr}.  

Measurements of EDMs (denoted $d$) are also sensitive to CP violation in and beyond the SM~\cite{Pospelov:2005pr}.  No EDM has yet been observed, but many future experiments await~\cite{Dzuba:2010dy}.
Currently, the most significant EDM bounds are for the neutron~\cite{Baker:2006ts}, atomic mercury ($^{199}$Hg)~\cite{Griffith:2009zz}, atomic thallium ($^{205}$Tl)~\cite{Regan:2002ta}, and recently molecular YbF~\cite{Hudson:2011zz}.  These null results provide important constraints on CP violation in the SM due to the $\theta_{\textrm{QCD}}$ phase associated with the strong interaction (present limits on $d_n$ require $\theta_{\textrm{QCD}} < 10^{-10}$~\cite{Pospelov:1999ha}), and on CP violation beyond the SM, such as in the Minimal Supersymmetric Standard Model (MSSM)~\cite{Ellis:2008zy,Li:2010ax}. On the other hand, these observables are rather insensitive to the KM phase, requiring many orders of magnitude increases in sensitivities (see Ref.~\cite{Pospelov:2005pr} and references therein).

In this work, we compare $D$ vs EDMs (in particular, $d_n$ and $d_{\textrm{Hg}}$) as probes of CP violation beyond the SM.  
For a given model, any CP-odd phase contributing to $D$ generates an ``irreducible'' EDM that can only be avoided by fine-tuned cancellations with other phases in the model.  We compute the resulting bounds on $D$ from EDMs in several new physics models:
left-right symmetric models~\cite{LRsym}, MSSM with R-parity violation~\cite{Barbier:2004ez}, models with exotic fermions~\cite{Langacker:1988ur}, and leptoquark (LQ) models~\cite{Buchmuller:1986zs}.  Most of these scenarios, and the resulting constraints from EDMs, have been studied previously~\cite{Herczeg:1982ij, Herczeg:2001vk, Herczeg:2005yf}.  
Here, we provide several improvements:
\begin{itemize}
\item We take into account recent improved computations of $d_n$~\cite{An:2009zh} and $d_{\textrm{Hg}}$~\cite{Ban:2010ea}.
Large uncertainties in the sensitivity of $d_{\textrm{Hg}}$ to the CP-odd isovector pion-nucleon coupling~\cite{Ban:2010ea} have weakened this constraint, and the $d_n$ bound currently provides the strongest limit on $D_t$.
\item In the literature, LQ contributions to $D_t$ are regarded as being safe from EDM constraints~\cite{Herczeg:2001vk,Herczeg:2005yf}.  We argue that $D_t$ is in fact more constrained than previously thought.  We also study implications for $D$ from LQ searches at hadron colliders.
\item We compute for the first time $D_t$ in the R-parity violating MSSM (with baryon-number violation), arising at one-loop order.
\item We provide a (partially) model-independent analysis that applies to all the aforementioned models except LQs, for which the current limit on $d_n$ implies $D_t < 3 \times 10^{-7}$.
\end{itemize}
We emphasize that $D$ is much cleaner theoretically than the EDMs constraining it, which rely on hadronic and nuclear computations.  Moreover, any realistic model may contain many different CP-odd phases, to which $D_t$ and EDMs are sensitive to different linear combinations.  The bounds we derive may be negated if there exist accidental cancellations between phases entering EDMs, and we neglect this possibility in our analysis.

Our work is organized as follows.  In Sec.~\ref{sec:main}, we review CP violation in $\beta$-decay.  We also summarize theoretical computations of neutron, mercury, and deuteron EDMs from underlying CP-violating operators most relevant for constraining $D_t$.
In Secs.~\ref{sec:ops} and \ref{sec:models}, we study constraints on $D_t$ from EDM bounds in several scenarios beyond the SM, focusing in particular on LQ models.  We present our conclusions in Sec.~\ref{sec:conclude}.

\section{CP-violating observables}
\label{sec:main}

\subsection{Beta decay}

The most general set of $\beta$-decay interactions can be parametrized at the quark level by an effective Lagrangian~\cite{Deutsch:1996qm}
\be
\label{eq:ham}
\mathscr{L}_{\beta} = - \frac{4 G_F V_{ud}}{\sqrt{2}} \sum_{\alpha,\beta,\gamma} a^\gamma_{\alpha\beta} \, \bar{e}_\alpha \Gamma^\gamma \nu_e \, \bar{u}  \Gamma_\gamma d_\beta + \; \textrm{h.c.}
\ee
The chiralities $(L,R)$ of the electron and down quark are labeled by $\alpha, \beta$.  The index $\gamma = S,V,T$ labels whether the interaction is scalar $(\Gamma^S\equiv 1)$, vector $(\Gamma^V \equiv \gamma^\mu)$, or tensor $(\Gamma^T \equiv \sigma^{\mu\nu}/\sqrt{2})$.  
CP invariance is preserved in $\beta$-decay if all ten complex coefficients
\be
a_{LL}^S, \, a_{LR}^S, \, a_{RL}^S, \, a_{RR}^S , \, a_{LL}^V, \, a_{LR}^V, \, a_{RL}^V, \, a_{RR}^V, \, a_{LR}^T, \, a_{RL}^T 
\ee
have a common phase ($a_{LL}^T, \, a_{RR}^T$ terms are identically zero).  
At leading order in the SM, all parameters vanish except $a_{LL}^V\!=\! 1$.  SM radiative corrections and new physics contributions to $a_{LL}^V$ can play an important role in the extraction of $V_{ud}$ (see, e.g., Refs.~\cite{Kurylov:2001zx}), but for CP-violating 
observables they can be neglected as subleading effects.  
We also hereafter set $V_{ud} = 1$; correlations between $D$ and EDMs depend on $|V_{ud}|$, but the 
$\mathcal{O}(\textrm{few}\,\%)$ deviation from $|V_{ud}| = 1$ is irrelevant compared to other theoretical uncertainties.  We neglect possible flavor constraints by considering only couplings between first generation fermions.  Lastly, we assume that $\beta$-decay processes involve a single neutrino flavor eigenstate $\nu_e$, and we allow for both $L,R$ chiralities.  Coefficients involving (sterile) 
right-handed neutrinos are only relevant provided these states are kinematically allowed in $\beta$-decay.\footnote{Sterile neutrinos with eV-scale mass 
have been studied recently in connection with various neutrino anomalies 
(see, e.g., Refs.~\cite{Kopp:2011qd}), and important constraints are provided by cosmology~\cite{Hamann:2011ge}. We do not attempt to accommodate these issues here.}   

In terms of the parametrization in Eq.~\eqref{eq:ham}, $D$ is given by~\cite{Jackson:1957zz}
\be \label{Deq}
D_t =  \kappa\, \textrm{Im}\left(a_{LR}^{V}a_{LL}^{V*} + a_{RL}^{V} a_{RR}^{V*} \right) 
+ \kappa\,\frac{g_S g_T}{g_V g_A} \,  \textrm{Im}\left(a_{L+}^{S} a_{LR}^{T*} + a_{R+}^{S} a_{RL}^{T*} \right) 
\ee
where $a_{L+}^{S} \equiv (a_{LL}^{S} + a_{LR}^{S})$ and $a_{R+}^{S} \equiv (a_{RL}^{S} + a_{RR}^{S})$.
For initial (final) state nucleus of spin $J$ ($J^\prime$), the coefficient $\kappa$ is 
\begin{align} \label{kappa}
\kappa \,\equiv\, \frac{ 4 g_V g_A M_{\textrm F} M_{\textrm{GT}}  }{ g_V^2 M_{\textrm F}^2 + g_A^2  M_{\textrm{GT}}^2} \, \sqrt{\frac{J}{J+1}} \, \delta_{JJ^\prime} \; \simeq \; \left\{ \ba{cl} 0.87 & \; \textrm{for} \; n \\ -1.03 & \; \textrm{for} \;  ^{19}\textrm{Ne} \ea \right. \;  , 
\end{align}
where $g_V = 1$, $g_A \approx 1.27$~\cite{Nakamura:2010zzi}, and $M_{\textrm F} \, (M_{\textrm{GT}})$ is the Fermi (Gamow-Teller) matrix element.  Scalar and tensor form factors $g_{S,T}$, originally estimated in Ref.~\cite{Adler:1975he}, have been computed using lattice techniques (see Ref.~\cite{Bhattacharya:2011qm} and references therein).
In this work, we neglect the scalar-tensor term in Eq.~\eqref{Deq}.  The $R$ coefficient, corresponding to the T-odd $\beta$-decay correlation $\langle \mathbf J \rangle \cdot \boldsymbol{\sigma}_e \times \mathbf p_e$ where 
$\boldsymbol{\sigma}_e$ is $e^\pm$ polarization,
has greater sensitivity to scalar- and tensor-type CP violation~\cite{Deutsch:1996qm, Herczeg:2001vk}.  Moreover, these couplings are correlated with CP-odd tensor and scalar electron-nucleon couplings, which are strongly constrained by $^{199}\textrm{Hg}$~\cite{Griffith:2009zz} and $^{205}\textrm{Tl}$~\cite{Regan:2002ta} EDM bounds, 
respectively~\cite{MP:1991,Khriplovich:1997ga,Ginges:2003qt,Sahoo:2008,Dzuba:2009mw}.

\subsection{Electric Dipole Moments}

\begin{figure}[t]
\begin{center}
%\vspace{1cm}
        \includegraphics[scale=1]{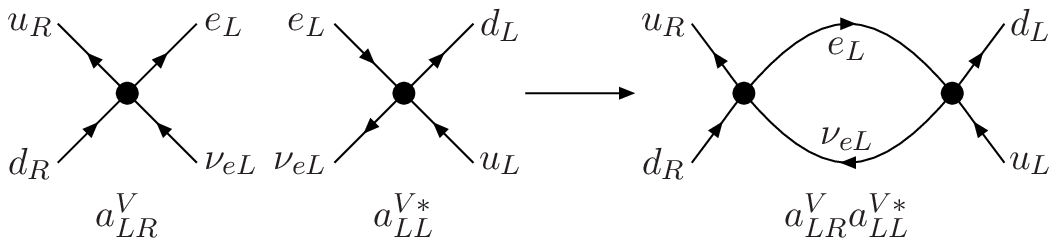} 
        \includegraphics[scale=1]{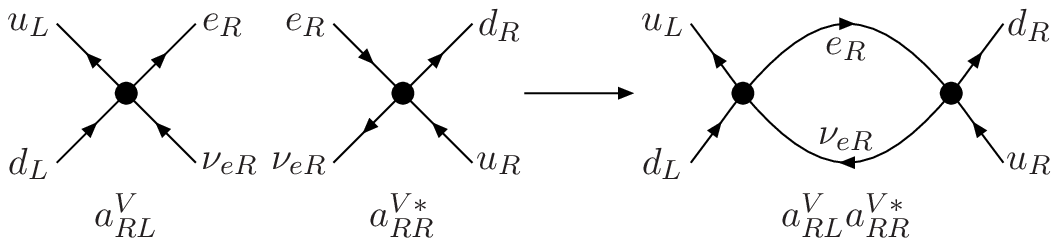} 
\end{center}
\caption{\it CP violation entering $D_t = \kappa \textrm{Im}(a_{LR}^V a_{LL}^{V*} + a_{RL}^V a_{RR}^{V*})$ automatically generates the four-quark operator $\mathcal{O}_{LR}\equiv i  (\bar{u}_L \gamma^\mu d_L \, \bar{d}_R \gamma_\mu u_R - \bar{d}_L \gamma^\mu u_L \, \bar{u}_R \gamma_\mu d_R)$, which contributes to neutron, mercury, and deuteron EDMs.\label{Dfig}
}
\end{figure}

EDM searches are sensitive to a wide class of CP-violating operators that can arise beyond the SM: 
CP-odd quark and lepton dipole moments, Weinberg's three-gluon operator~\cite{Weinberg:1989dx}, and four-fermion operators.  Here, the most relevant one is a CP-odd four-quark operator $\mathcal{O}_{LR}$, given by
\be
\mathscr{L}_{\textrm{eff}} = - \frac{4G_F}{\sqrt{2}} \, k_{LR} \,\mathcal{O}_{LR}  \, ,\quad
\mathcal{O}_{LR} \equiv i  (\bar{u}_L \gamma^\mu d_L \, \bar{d}_R \gamma_\mu u_R - \bar{d}_L \gamma^\mu u_L \, \bar{u}_R \gamma_\mu d_R) \label{Lqq} 
\ee
where $k_{LR}$ is the operator coefficient (normalized to $4G_F/\sqrt{2}$).
Within the context of left-right symmetric models, this effective interaction arises from CP-violating $W$-$W^\prime$ mixing and has been studied previously~\cite{Khatsimovsky:1987fr,He:1989xj,He:1992db,Xu:2009nt}. 
We show in Fig.~\ref{Dfig} that, by connecting the leptonic legs in a one-loop diagram, the same interference terms $a_{LR}^V a_{LL}^{V*}$ and $a_{RL}^V a_{RR}^{V*}$ contributing to $D_t$ also generate $\mathcal{O}_{LR}$.  Moreover, this diagram does not involve any chirality-changing mass insertions, and therefore is not suppressed by any light fermion masses.   Other CP-odd operators (e.g., quark EDMs) also arise from new physics entering $D_t$, but are suppressed by light masses and will not be considered here.

%---------------------------------------------------------
%---------------------------------------------------------

The most significant EDM constraints on $\mathcal{O}_{LR}$ are for the neutron~\cite{Baker:2006ts} and mercury atom~\cite{Griffith:2009zz}:
\begin{align}
|d_{n}| < 2.9 \times 10^{-26} \; e \, \textrm{cm} \quad (90\% \; \textrm{CL})  \; , \quad
|d_{\textrm{Hg}}| < 3.1 \times 10^{-29} \; e \, \textrm{cm} \quad (95\% \; \textrm{CL})\; . 
\end{align}
Future measurements of the deuteron EDM $d_D$, expected at the level of \mbox{$10^{-27} \; e$ cm} or better~\cite{Semertzidis:2009zza}, will also provide important constraints on $\mathcal{O}_{LR}$.

Ref.~\cite{An:2009zh} has performed a systematic computation of $d_n$ from CP-odd four-fermion operators, using a combination of chiral effective theory techniques and quark model estimates for the hadronic matrix elements.  
Using their results, we take
\be
d_{n} = - 1 \times 10^{-19} \, k_{LR} \; e \, \textrm{cm}  \; ,
\ee
with an $\mathcal{O}(1)$ uncertainty on the numerical prefactor~\cite{An:2009zh}.\footnote{This value is consistent with a naive estimate $d_n \sim e M_{QCD}/\Lambda^2$, where  $M_{QCD} \sim 1$ GeV is the QCD scale and $\Lambda$ is the scale of CP violation.  Taking $\Lambda^{-2} \sim G_F k_{LR}$, we have $|d_n| \sim 2 |k_{LR}| \times 10^{-19} \; e \, \textrm{cm}$.  Also, it is useful to note $\mathcal{O}_{LR} = (\bar u u \bar{d} i \gamma_5 d - \bar{u} i \gamma_5 u \bar d d + 6 \bar{u}  t^a u \bar{d} i \gamma_5 t^a d - 6\bar{u} i \gamma_5  t^a u \bar{d}t^a d)/3$ using a Fierz transformation, 
where $t^a$ is the $SU(3)_c$ generator, to make contact with the notation of Ref.~\cite{An:2009zh}.}
Earlier results~\cite{Khatsimovsky:1987fr,He:1992db,Khatsymovsky:1992yg,Khriplovich:1997ga} are consistent at the order-of-magnitude level, but according to Ref.~\cite{An:2009zh} are not as reliable in that they take into account different subsets of the full set of contributions to $d_n$.  

Diamagnetic atoms (e.g., $^{199}$Hg) are also sensitive to $CP$-odd four-quark interactions.  
Interpretation of these measurements is a three step process (see, e.g., Ref.~\cite{Khriplovich:1997ga,Ginges:2003qt}).  First, atomic calculations relate the measured EDM to the nuclear Schiff moment $S$.
For the case of mercury, we take~\cite{Dzuba:2009kn}
\be
d_{\textrm{Hg}} = -2.6 \times 10^{-17} \, e \, \textrm{cm} \times \left( \frac{S_{\textrm{Hg}}}{e \, \textrm{fm}^3} \right) \; .
\ee
This numerical value (2.6) agrees with an earlier result (2.8) by two of those authors~\cite{Dzuba:2002kg}, while another recent computation found a larger value (5.1)~\cite{Latha:2009nq}. Second, the Schiff moment is computed in terms of $P,T$-odd nucleon-pion couplings, of which only the isovector coupling $\bar{g}_1$ is relevant since $\mathcal{O}_{LR}$ is isovector~\cite{Herczeg:1987}.
Previous nuclear computations found (keeping only $\bar{g}_1$ terms): $S_{\textrm{Hg}} = - 0.071 \, g \bar{g}_1 \; e \, \textrm{cm}^3$~\cite{deJesus:2005nb} and $S_{\textrm{Hg}} = - 0.055 \, g \bar{g}_1 \; e \, \textrm{cm}^3$~\cite{Dmitriev:2003kb}, where $g \approx 13.5$ is the (CP-even) pion-nucleon strong coupling.  However, a recent and improved computation by Ref.~\cite{Ban:2010ea} found that the $\bar{g}_1$ coefficient is very sensitive to the model-dependent nuclear potential inputs and may be suppressed by an order of magnitude (or more) and may have opposite sign compared to Refs.~\cite{deJesus:2005nb,Dmitriev:2003kb}.  
These nuclear physics uncertainties are crucial for constraining $D_t$ using $d_{\textrm{Hg}}$.  
In light of this unresolved issue, 
we take $|S_{\textrm{Hg}}| = 0.01 \, g |\bar g_{1}|\; e \, \textrm{fm}^3$, remaining agnostic as to the sign (see Ref.~\cite{Ellis:2011hp} for additional discussion). 
Third, following Ref.~\cite{Khatsimovsky:1987fr}, we conservatively take $\bar{g}_1 = 2 \times 10^{-6} \, k_{LR}$.  Ref.~\cite{He:1992jh} found a larger numerical prefactor by a factor of 7.  Putting all these pieces together, we take
\be
|d_{\textrm{Hg}}| = 7 \times 10^{-24} \, |k_{LR}| \; e \, \textrm{cm} \; ,
\ee
with an uncertainty at the order-of-magnitude level.

The deuteron EDM provides a much cleaner probe of $\bar{g}_1$ compared to $d_{\textrm{Hg}}$.
Following the recent computation of Ref.~\cite{deVries:2011an} (in good agreement with earlier results~\cite{Khriplovich:1999qr,Liu:2004tq,Korkin:2005bw}), we take
\be
|d_D| \approx 1.9 \times 10^{-14} \, |\bar{g}_1| \;  e \, \textrm{cm}  \approx 4.5 \times 10^{-20} \, |k_{LR}| \; e \, \textrm{cm}  \; ,
\ee
with $\mathcal{O}(20\!-\!30\%)$ uncertainty on the numerical factor (1.9)~\cite{Afnan:2010xd,deVries:2011an}.

\section{Model-independent bounds on $D$}
\label{sec:ops}

\begin{figure}[t]
\begin{center}
%\vspace{1cm}
        \includegraphics[scale=.9]{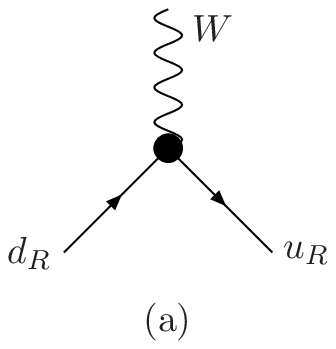} \qquad
        \includegraphics[scale=.9]{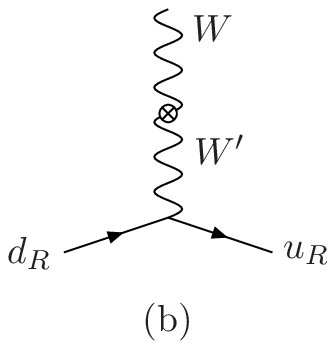} \qquad
        \includegraphics[scale=.9]{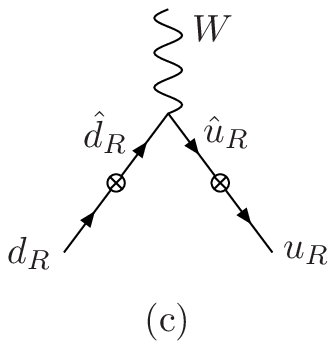} \qquad
        \includegraphics[scale=.9]{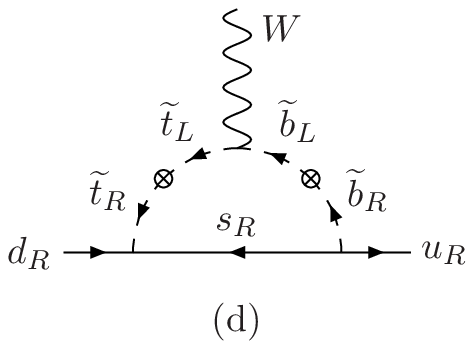} 
\end{center}
\caption{\it (a) Effective $\bar u_R \gamma^\mu d_R W^+_\mu$ vertex arising beyond the SM: e.g., 
(b) left-right symmetric model with $W$-$W^\prime$ mixing; (c) exotic quarks $\hat{u}_R$, $\hat{d}_R$ with non-standard $SU(2)_L \times U(1)_Y$ gauge couplings that mix with SM quarks $u_R$, $d_R$; (d) $R$-parity violating MSSM with baryon number violation and squark left-right mixing.  In each case, mixing insertions (involving the Higgs vev $v$) are denoted by $\otimes$.}
\label{dim6fig}
\end{figure}

New physics contributions to $\beta$-decay can be organized in terms of a hierarchy of non-renormalizable operators characterized by mass scale $\Lambda >G_F^{-1/2}$.  Naively, the leading contributions to $D_t$ will be those suppressed by the fewest powers of $(G_F \Lambda^2)^{-1}$: namely, from dimension-six operators contributing to $a_{LR}^V$ that interfere with the SM amplitude $a_{LL}^V$.  There is only one such operator~\cite{Buchmuller:1985jz}:
\be
\mathscr{L}_{\textrm{dim} \, 6} \; = \; \frac{c}{\Lambda^2} \, \bar{u}_R \gamma^\mu d_R \, i H^T \epsilon D_\mu H \; + \; \textrm{h.c.} \; , \label{Dop6}
\ee
where $c$ is a complex coefficient.  $H$ is the Higgs doublet, $D_\mu$ is its covariant derivative, $\epsilon$ is the antisymmetric tensor ($\epsilon_{12} = 1$), and $^T$ denotes transpose acting on $SU(2)_L$ indices.  Setting the Higgs field equal to its vacuum expectation value, Eq.~\eqref{Dop6} generates a coupling of the $W$ boson to the right-handed charge current $\bar{u}_R \gamma^\mu d_R$, shown in Fig.~\ref{dim6fig}.  Integrating out the $W$ boson, we obtain (recall we set $V_{ud} = 1$)
\be
\mathscr{L}_{\textrm{dim} \, 6} \; = \; - \; \frac{c}{\Lambda^2} \left( \,\bar{u}_R \gamma^\mu d_R \, \bar{e}_L \gamma_\mu \nu_{eL} + \bar{u}_R \gamma^\mu d_R \, \bar{d}_L \gamma_\mu u_{L} \,\right) \; + \; \textrm{h.c.} \; 
\label{D2R}
\ee 
The operator of Eq.~\eqref{Dop6} generates at order $(G_F \Lambda^2)^{-1}$
contributions to both $a_{LR}^V$ and $k_{LR}$:
\be
\text{Im}(a_{LR}^V) = k_{LR} = \frac{ \textrm{Im}(c)}{2\sqrt{2} \, G_F \Lambda^2} \; .
\ee
For all models that contribute to $D_t$ via Eq.~\eqref{Dop6}, EDMs are correlated with $D_t$ in an otherwise model-independent way:
\begin{subequations} \label{dim6bounds}
\begin{align}
|d_n| &= 1 \times 10^{-19} \; e \, \textrm{cm} \times |D_t/\kappa| \\
|d_{\textrm{Hg}}| &= 7 \times 10^{-24} \; e \, \textrm{cm} \times |D_t/\kappa| \\
|d_D| &= 4.5 \times 10^{-20} \; e \, \textrm{cm} \times |D_t/\kappa| \; .
\end{align}
\end{subequations}
The current bound $|d_n| < 2.9 \times 10^{-26} \, e \, \textrm{cm}$~\cite{Baker:2006ts} implies $|D_t/\kappa| < 3 \times 10^{-7}$, far below present sensitivities.

%---------------------------------------------------------
%---------------------------------------------------------

This indirect limit on $D_t$ applies to the following models:
\begin{itemize}
\item Left-right symmetric models with a $W^\prime$ boson that mixes with the $W$ and couples to the right-handed quark charge current.
\item Models with exotic fermions with non-standard gauge quantum numbers, e.g., exotic $SU(2)_L$-doublet vector quarks $\hat{u}$ and $\hat{d}$ that mix with the usual $u$ and $d$ quarks.
\item $R$-parity violating (RPV) MSSM with baryon number violation, described below~\cite{Drees:2003dv}.
\end{itemize}
The relevant diagrams are shown in Fig.~\ref{dim6fig}.  The first two models were studied previously in connection with $D$ in Refs.~\cite{Herczeg:1982ij, Herczeg:2001vk, Herczeg:2005yf}, and we do not describe them here.  

The RPV MSSM is defined by adding to the MSSM superpotential gauge-invariant and renormalizable
terms that violate either baryon or lepton number 
(but not both, to avoid proton decay)~\cite{Barbier:2004ez}.  
Contributions to $D_t$ are generated by the baryon number-violating terms\footnote{Lepton number-violating terms have been studied previously in connection with the $R$ coefficient~\cite{Yamanaka:2009hi}.} 
\be
W_{RPV} = \lambda^{\prime\prime}_{ijk} \, U^c_i D^c_j D^c_k \; ,
\ee
where $U^c_i$, $D^c_j$ are superfields corresponding to the (charge-conjugate) $u_R^i$ and $d_R^j$ quarks of generation $i,j$, respectively.   Shown in Fig.~\ref{dim6fig}, the leading contributions to $D_t$ arise at one-loop from diagrams involving third generation squarks $\widetilde t_{L,R}$ and $\widetilde b_{L,R}$.  This contribution relies on mixing between gauge eigenstates, described by (see, e.g., Ref.~\cite{Martin:1997ns})
\be \label{LRmix}
\mathscr{L}_{\textrm{mix}} 
= - m_t\, (A_t^* \sin\beta + \mu \cos\beta)\, \widetilde t_L^\dagger \widetilde t_R
- m_b \, (A_b^* \cos\beta + \mu \sin\beta) \, \widetilde b_L^\dagger \widetilde b_R + \; \textrm{h.c.}
\ee
where $\tan\beta$ is the ratio between up- and down-type Higgs vacuum expectation values, $A_{t,b}$ and $\mu$ are MSSM mass parameters, and $m_{t}$ ($m_{b}$) is the top (bottom) quark mass.  
For $\tan\beta \gg 1$, we have
\be
a_{LR}^V = \frac{\lambda^{\prime\prime}_{123}\lambda^{\prime\prime*}_{312} V_{tb}\, m_t m_b \tan\beta\,\mu A_t}{24\pi^2 \, m_{\widetilde q}^4} \; . \label{RPVaLR}
\ee
assuming degenerate squarks with mass $m_{\widetilde q}$ and treating Eq.~\eqref{LRmix} perturbatively by mass insertion.  Bounds on $n$-$\bar{n}$ oscillations constrain 
$|\lambda_{312}| \lesssim 10^{-2}$ if all squarks have mass $m_{\widetilde q} = 200$ GeV~\cite{Chang:1996sw}, but this bound is relaxed if only third generation squarks are light; $|\lambda_{123}|$ is unconstrained~\cite{Barbier:2004ez}.
In principle Eq.~\eqref{RPVaLR} can be as large as $\mathcal{O}(10^{-3})$ for $m_{\widetilde q}, A_t, \mu \sim 200$ GeV, $\lambda^{\prime\prime}\sim 1$, and $\tan\beta \sim 50$ (perturbativity of the bottom Yukawa coupling requires $\tan\beta \lesssim 60$).  However, the neutron EDM bound constrains $\textrm{Im}(a_{LR}^V) < 3 \times 10^{-7}$, as per our previous discussion.

Ref.~\cite{Godbole:1999ye} previously studied the RPV MSSM in connection with EDMs, focusing on contributions from quark and electron CP-odd dipole moments arising at two-loop.  For the combination of RPV couplings 
$\lambda^{\prime\prime}$ in Eq.~\eqref{RPVaLR} entering $D_t$, quark EDM and chromo-EDM operators are suppressed by $m_{u,d}$.  Here, the CP-odd four-quark operator gives a much stronger bound.

%%%%%%%%%%%%%%%%%%%%%%%%%%%%%%%%

\section{Leptoquark models}
\label{sec:models}

Leptoquarks (LQs), fractionally-charged colored states carrying baryon and lepton number, arise in many extensions of the SM, e.g., grand unification~\cite{Georgi:1974sy} and compositeness~\cite{Schrempp:1984nj}.  Here, we consider a phenomenological model of LQs coupled to first generation quarks and leptons~\cite{Buchmuller:1986zs}.  LQ models have a rich phenomenology for $\beta$-decay, potentially giving large contributions to $D$ and other observables through tree-level processes~\cite{Herczeg:2001vk}.  

In the literature, LQ models have been regarded as an ``EDM safe'' source of CP violation that might generate $D$ as large as present experimental limits, without fine tuning~\cite{Herczeg:2001vk}.  These previously considered models (dubbed the ``usual scenarios'') rely on LQ mixing to generate a dimension-eight operator contributing to $a_{LR}^V$ at tree-level, which interferes with the SM amplitude $a_{LL}^V$~\cite{Herczeg:2001vk}.  In addition, scenarios involving LQs coupled to right-handed neutrinos can also generate $D_t$ via the interference of two new physics amplitudes $a_{RL}^V$ and $a_{RR}^V$.  

In this section, we study in detail these cases (i.e., with or without right-handed neutrinos).  We show that radiative corrections generate contributions to EDMs 
(in the spirit of Refs.~\cite{Khriplovich:1990ef,Kurylov:2000ub}) sensitive to the same phases entering $D_t$.  In both cases, the resulting bounds from the neutron EDM are stronger than the direct experimental limit.

\subsection{Usual LQ scenarios: no right-handed neutrinos}

%---------------------------------------------------------
\begin{figure}[t]
\begin{center}
        \includegraphics[scale=1]{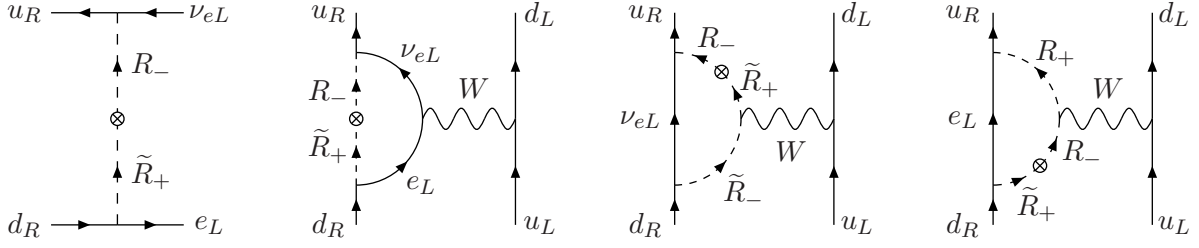} 
\end{center}
\caption{\it Scalar LQ case: tree-level exchange generates $\beta$-decay amplitude $a_{LR}^V$ (left), while 
$\mathcal{O}_{LR}$ is generated by one-loop vertex corrections (right), contributing to EDMs $d_n$, $d_{\textrm{Hg}}$, $d_D$.
Diagrams are shown in weak-eigenstate LQ basis to illustrate that the same CP-violating phases from LQ-mixing (denoted $\otimes$) and couplings enter both $D_t$ and EDMs.
}
\label{LQfigL1}
\end{figure}
%
%-----------------------------------------------------

There are two cases to consider: scalar and vector LQ exchange, both considered previously in Ref.~\cite{Herczeg:2001vk}.  
Since both cases are similar, we treat them simultaneously.  The relevant LQs are
\begin{subequations}\begin{align}
&\textrm{scalar case:} & R &= \left( \ba{c} R_+ \\ R_- \ea \right) \sim ( 3, 2, 7/6) &
\widetilde R &= \left( \ba{c} \widetilde R_+ \\ \widetilde R_- \ea \right) \sim (3, 2, 1/6) \\
&\textrm{vector case:} & V &= \left( \ba{c} V_+ \\ V_- \ea \right) \sim (\bar 3, 2, 5/6)&
\widetilde V &= \left( \ba{c} \widetilde V_+ \\ \widetilde V_- \ea \right) \sim (\bar 3, 2, -1/6)
\end{align}
\end{subequations}
where $\pm$ states are weak isospin components, and the $SU(3)_C \times SU(2)_L \times U(1)_Y$ quantum numbers are given in  parentheses.\footnote{We follow the notation of Ref.~\cite{Buchmuller:1986zs} for LQ states, except we omit an additional subscript identifying the $SU(2)_L$ representation.}  In both cases, the most general renormalizable interactions to first generation fermions (including $\nu_{eR}$) are
\begin{subequations} \label{LQLint}
\begin{align}
\textrm{scalar:}\; & 
\mathscr{L}_{\textrm{int}} = h_L \, \bar u_R L_L^T \epsilon R + h_R \, \bar Q_L e_R R + \tilde{h}_L \, \bar d_R L_L^T \epsilon\widetilde R  + \tilde{h}_R \, \bar Q_L \nu_{eR} \widetilde R + \textrm{h.c.} \\
\textrm{vector:}\; & 
\mathscr{L}_{\textrm{int}} = g_L \, \bar d_R^c \gamma^\mu L_L^T \epsilon V_\mu + g_R \, \bar Q_L^c \gamma^\mu e_R \,\epsilon  V_\mu + \tilde{g}_L \, \bar u_R^c \gamma^\mu L_L^T \epsilon \widetilde V_\mu  + \tilde{g}_R \, \bar Q_L^c \gamma^\mu \nu_{eR} \,\epsilon  \widetilde V_\mu + \textrm{h.c.} 
\end{align}
\end{subequations}
with quark and lepton doublets $Q_L = (u_L, d_L)$ and $L_L = (\nu_{eL}, e_L)$.  Here, $h_{L,R}$, $\tilde{h}_{L,R}$, $g_{L,R}$, $\tilde{g}_{L,R}$ are couplings (with $L,R$ denoting lepton chirality).  The presence of both $L,R$-type couplings will lead to lepton universality violation in $\pi^+ \to e^+ \nu$; to avoid this constraint, we set  $R$-type couplings to zero~\cite{Buchmuller:1986zs}.  The relevant mass terms are
\begin{subequations}\label{LQmix}\begin{align}
&\textrm{scalar:} & 
- \mathscr{L}_{\textrm{mass}} &= m_R^2 R^\dagger R + m_{\widetilde R}^2 \widetilde R^\dagger \widetilde R
+ \left( \lambda_R (R^\dagger H)(\widetilde R H) + \textrm{h.c.} \right) \\
&\textrm{vector:} & 
\mathscr{L}_{\textrm{mass}} &= m_V^2 V_\mu^\dagger V^\mu + m_{\widetilde V}^2 \widetilde V_\mu^\dagger \widetilde V^\mu
+ \left( \lambda_V (V_\mu^\dagger H)(\widetilde V^\mu H) + \textrm{h.c.} \right) 
\end{align}
\end{subequations}
Through electroweak symmetry breaking, the quartic interactions (with couplings $\lambda_{R,V}$) give rise to $R_-$-$\widetilde R_+$ mixing and $V_-$-$\widetilde V_+$ mixing by generating off-diagonal mass terms proportional to $\lambda_{R,V} v^2$, where $v \equiv \langle H^0 \rangle$.  Diagonalizing the $R_-$-$\widetilde R_+$ and $V_-$-$\widetilde V_+$ mass matrices, we can express the mass eigenstates, denoted $\mathcal{R}_{1,2}$ and $\mathcal{V}_{1,2}$, as 
\begin{subequations}\begin{align}
&\textrm{scalar:}  & \mathcal{R}_1 &\equiv \cos\theta_R \,R_- + \sin\theta_R \, e^{i \phi_R} \widetilde R_+ \, , &
\mathcal{R}_2 &\equiv \cos\theta_R \, \widetilde R_+ - \sin\theta_R \, e^{-i \phi_R} R_-  \\
&\textrm{vector:} & \mathcal{V}_1 &\equiv \cos\theta_V \,V_- + \sin\theta_V \, e^{i \phi_V} \widetilde V_+ \, , &
\mathcal{V}_2  &\equiv \cos\theta_V \, \widetilde V_+ - \sin\theta_V \, e^{-i \phi} V_- \end{align}
\end{subequations}
with mixing angles $\theta_{R,V}$ and mass eigenvalues given by
\begin{subequations} \label{mixparam}
\begin{align}
&\textrm{scalar:} & \tan2\theta_R &= \frac{2|\lambda_R|v^2}{m_R^2 - m_{\widetilde R}^2 } ,
&  m_{\mathcal{R}_{1,2}}^2 &= \frac{1}{2} \left(m^2_{R} + m_{\widetilde R}^2 \pm \sqrt{ (m_{R}^2 - m_{\widetilde R}^2)^2 + 4 |\lambda_{R}|^2 v^4 }\right) \\
&\textrm{vector:}& \tan2\theta_V &= \frac{2|\lambda_V|v^2}{m_V^2 - m_{\widetilde V}^2 } ,  
& m_{\mathcal{V}_{1,2}}^2 &= \frac{1}{2} \left(m^2_{V} + m_{\widetilde V}^2 \pm \sqrt{ (m_{V}^2 - m_{\widetilde V}^2)^2 + 4 |\lambda_{V}|^2 v^4 }\right) 
\end{align}
\end{subequations}
and phases $\phi_{R,V} = \arg(\lambda_{R,V})$, defined such that $m_{\mathcal{R},\mathcal{V}_1}^2 < m_{\mathcal{R},\mathcal{V}_2}^2$.
The remaining (unmixed) LQ states $R_+$, $V_+$ and $\widetilde R_-$, $\widetilde V_-$ have masses $m_{R,V}$ and $m_{\widetilde R, \widetilde V}$, respectively.

%------------------------------------------------------

\begin{figure}[t]
\begin{center}
        \includegraphics[scale=1]{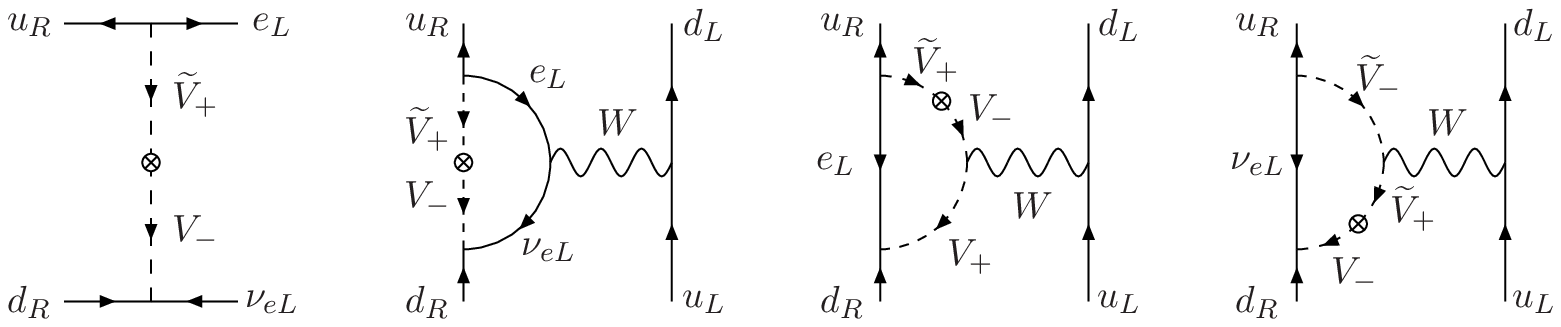} 
\end{center}
\caption{\it Vector LQ case: Tree-level exchange generates $\beta$-decay amplitude $a_{LR}^V$ (left), while 
$\mathcal{O}_{LR}$ is generated by one-loop vertex corrections (right), contributing to EDMs $d_n$, $d_{\textrm{Hg}}$, $d_D$.
Diagrams are shown in weak-eigenstate LQ basis to illustrate that the same CP-violating phases from LQ-mixing (denoted $\otimes$) and couplings enter both $D_t$ and EDMs.
}
\label{LQfigL2}
\end{figure}

%---------------------------------------------------------

For $\beta$-decay, this model gives $D_t = \kappa \, \textrm{Im}(a_{LR}^V)$, where \cite{Herczeg:2001vk}
\begin{subequations} \label{aLRV}
\begin{align}
&\textrm{scalar case:} \quad  a_{LR}^V = \frac{h_L \tilde h_L^* \sin 2\theta_R\, e^{i \phi_R} }{8\sqrt{2} G_F } 
\left( \frac{1}{m_{\mathcal{R}_1}^2} -\frac{1}{ m_{\mathcal{R}_2}^2} \right)  \\
&\textrm{vector case:} \quad a_{LR}^V = \frac{g_L \tilde g_L^* \sin 2\theta_V\, e^{i \phi_V} }{4\sqrt{2} G_F } 
\left( \frac{1}{m_{\mathcal{V}_1}^{2}} -\frac{1}{m_{\mathcal{V}_2}^{2}} \right) \; .
\end{align} 
\end{subequations}
The relevant Feynman diagrams are shown in Figs.~\ref{LQfigL1} and~\ref{LQfigL2}.

Next, we consider implications for EDMs.  Radiative corrections involving the $W$ boson, shown in 
Figs.~\ref{LQfigL1} and \ref{LQfigL2}, generate the CP-odd four-quark operator $\mathcal{O}_{LR}$ given 
in Eq.~\eqref{Lqq} which contributes to $d_n$ and $d_{\textrm{Hg}}$.  The resulting coefficient $k_{LR}$ is proportional to the same CP-violating phases in Eqs.~\eqref{aLRV} entering $D$.  For each case, we find
\be \label{DkLR}
\textrm{scalar:} \;\; k_{LR} = \frac{8 G_F m_{\mathcal{R}_1}^2}{\sqrt{2} (4\pi)^2} \, F_{R} \, \textrm{Im}(a_{LR}^V) \, , \quad
\textrm{vector:} \;\; k_{LR} = - \frac{8 G_F m_{\mathcal{V}_1}^2}{\sqrt{2} (4\pi)^2} \,  F_{V} \, \textrm{Im}(a_{LR}^V) \; .
\ee
The loop functions $F_{R,V}$ are given by
\begin{subequations}
\begin{align}
F_R &\equiv \frac{m_{\mathcal{R}_2}^2}{2(m_{\mathcal{R}_1}^2 - m_{\mathcal{R}_2}^2)} \left( f(m_{\mathcal{R}_1}^2,m_{\mathcal{R}_2}^2,m_R^2) + f(m_{\mathcal{R}_1}^2,m_{\mathcal{R}_2}^2,m_{\widetilde R}^2) + f(m^2_{\mathcal{R}_1},m^2_{\mathcal{R}_2},0) \right) \\
F_V &\equiv \frac{m_{\mathcal{V}_2}^2}{2(m_{\mathcal{V}_1}^2 - m_{\mathcal{V}_2}^2)} \left( 3 f(m_{\mathcal{V}_1}^2,m_{\mathcal{V}_2}^2,m_V^2) + 3 f(m_{\mathcal{V}_1}^2,m_{\mathcal{V}_2}^2,m_{\widetilde V}^2) -  f(m^2_{\mathcal{V}_1},m^2_{\mathcal{V}_2},0) \right)
\end{align}
\end{subequations}
where
\be
f(m_1^2,m_2^2,m_3^2) \equiv \frac{m_1^2 m_2^2 \log(m_1^2/m_2^2) + m_2^2 m_3^2 \log(m_2^2/m_3^2)+ m_3^2 m_1^2 \log (m_3^2/m_1^2)}{(m_1^2 - m_3^2)(m_2^2 - m_3^2)} \; .
\ee
Defined in this way, we have $F_{R,V} \ge 1$, with equality in the limit $m_{R,V}^2 = m_{\widetilde R,\widetilde V}^2 \gg \lambda_{R,V} v^2$.
Eq.~\eqref{DkLR} provides the leading contributions to EDMs from CP violation entering $D$; there is no suppression by light quark masses $m_{u,d}^2$ as previously argued~\cite{Herczeg:2001vk}.  Using the results of Sec.~\ref{sec:main}, we have (both cases give the same numerical factors)
\begin{subequations} \label{LQLbound}
\begin{align}
|d_n| & > 4 \times 10^{-21} \; e \, \textrm{cm} \times \left|{ D_t}/{\kappa} \right| \; \left( \frac{m_{LQ}}{300 \; \textrm{GeV}} \right)^2  \\
|d_{\textrm{Hg}}| & > 3 \times 10^{-25}  \; e \, \textrm{cm}  \times \left|{ D_t}/{\kappa} \right| \; \left( \frac{m_{LQ}}{300 \; \textrm{GeV}} \right)^2  \\
|d_D| & > 1.7 \times 10^{-21}  \; e \, \textrm{cm}  \times \left|{ D_t}/{\kappa} \right| \; \left( \frac{m_{LQ}}{300 \; \textrm{GeV}} \right)^2
\end{align}
\end{subequations}
where $m_{LQ} = m_{\mathcal{R}_1}$ ($m_{\mathcal{V}_1}$) corresponds to the lightest LQ state entering $\beta$-decay for the scalar (vector) LQ case.  ($\kappa \approx 0.87, \, -1.03$ for $n, \, ^{19}\textrm{Ne}$, respectively.) 

Recent searches at hadron colliders~\cite{Khachatryan:2010mp, Aad:2011uv, Abazov:2011qj,D0:2009gf}
provide constraints on the mass of the lightest LQ $(\mathcal{R}_1,\mathcal{V}_1)$ involved in $\beta$-decay.  These bounds depend on the branching ratio $\beta_e \equiv \textrm{BR}(LQ \to j e) = 1 - \textrm{BR}({LQ} \to j \nu) =\sin^2\theta_{R,V}$, where $j$ is a jet. For the scalar case, the strongest limits have been obtained at the Large Hadron Collider by combining $jjee$ and $jje\nu$ channels~\cite{Khachatryan:2010mp,Aad:2011uv}:  
\be \label{cols}
\textrm{scalar case:} \quad
m_{\mathcal{R}_1} \; > \; \left\{ \ba{ll} 340 \; \textrm{GeV} & \;  \textrm{(CMS)} \\
319 \; \textrm{GeV} & \; \textrm{(ATLAS)} \ea \right. \quad  (\beta_e > 0.5)
\ee
with stronger limits (384 and 376 GeV, respectively) for $\beta_e \to 1$.  
Additionally, recent ATLAS searches for jets with missing energy from squark pair production, within a simplified SUSY context~\cite{Aad:2011ib}, apply to $jj\nu\nu$ final states from $\mathcal{R}_1$ pair production.  To translate the SUSY model into our framework, one must rescale the SUSY cross section by a factor $(1-\beta_e)^2/4$ and take the gluino to be massive.\footnote{The factor 4 counts the number of first and second generation squarks in the simplified SUSY model considered in Ref.~\cite{Aad:2011ib}.}  The resulting limits are $m_{\mathcal{R}_1} \gtrsim 500$ GeV, for $\beta_e < 0.5$, with stronger bounds in the limit $\beta_e \to 0$.  In the vector case, the D0 collaboration has obtained~\cite{D0:2009gf,Abazov:2001mx}
\be \label{colv}
\textrm{vector case:} \quad m_{\mathcal V_1} \; > \; 
\left\{ \ba{cc} 302 \; \textrm{GeV} &\;\; (jjee + jje\nu, \; \beta_e > 0.1) \\
144 \; \textrm{GeV} &\;\; (jj\nu\nu, \; \beta_e < 0.1) \ea \right.
\ee
with stronger bounds for $\beta_e \to 1$ or for different choices of anomalous gluon-LQ couplings considered therein.  Within the context of our model, for $\beta_e = \sin^2\theta_V < 0.1$, the lightest vector LQ $\mathcal{V}_1 \approx V_-$ is nearly degenerate with $V_+$ ($m_{\mathcal{V}_1} \approx m_V$).  Since $\textrm{BR}(V_+ \to je)=1$, we have $m_{V} > 367$~\cite{D0:2009gf}, and therefore $m_{\mathcal{V}_1}$ is constrained indirectly to be much heavier than 144 GeV. 
Additional constraints have been obtained by the H1 collaboration at HERA.
\footnote{To translate between the notation used here and that in Ref.~\cite{Collaboration:2011qa}, we note $S^L_{1/2} \equiv R$, $\widetilde S^L_{1/2} \equiv \widetilde R$, $V^L_{1/2} \equiv V$, and $\widetilde V^L_{1/2} \equiv \widetilde V$.}  These limits depend on the LQ-fermion couplings, and provide stronger bounds than those from hadron colliders if the relevant $e$-$q$-LQ coupling is larger than $\sim\textrm{few} \times 10^{-1}$~\cite{Collaboration:2011qa}.

Given the current limit $|d_n| < 2.9 \times 10^{-26} \; e \, \textrm{cm}$~\cite{Baker:2006ts}, and conservatively taking $m_{LQ} > 300$ GeV, we conclude that $|D_t/\kappa| < 7 \times 10^{-6}$.  CP violation in LQ models cannot saturate present experimental sensitivities in $D_n$ --- unless the hadronic uncertainties associated with the $d_n$ computation of Ref.~\cite{An:2009zh} are larger by an order of magnitude, or unless there is a cancellation with other CP-odd phases contributing to $d_n$ to $\sim 10\%$ (or a combination thereof).  On the other hand, the mercury EDM does not strongly constrain $D_t$ in this model, especially in light of its large hadronic uncertainties, although this situation may change with future refinements in the nuclear computations.

\subsection{LQ scenarios with right-handed neutrinos}

%---------------------------------------------------------
\begin{figure}[t]
\begin{center}
        \includegraphics[scale=1]{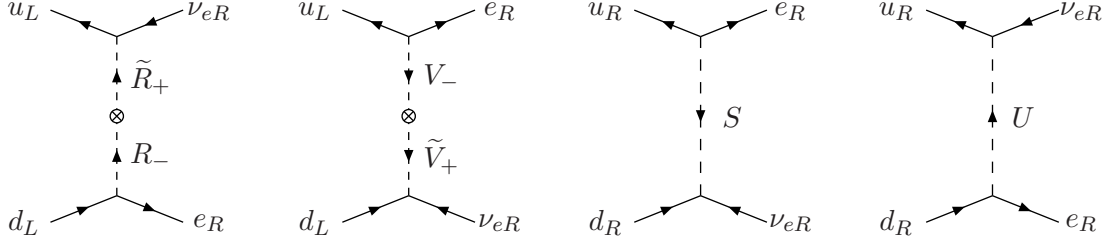} 
\caption{\it $D_t = \kappa \, \textrm{Im}(a_{RL}^V a_{RR}^{V*})$ is generated by LQ couplings involving right-handed neutrinos, with $a_{RL}^V$ from $R,\widetilde R$- or $V,\widetilde V$-exchange (mixing denoted $\otimes$), and $a_{RR}^V$ from $S$- or $U$-exchange.
}
\label{LQfigR1}
\end{center}
\end{figure}
%%%%%%%%%%%%%%%%%%%%%%%%%%%%%%%%%%%%%%%%%%%%%%5

LQ models can contribute to $D_t$ through the interference between two new physics amplitudes involving right-handed neutrinos.  The relevant Feynman diagrams are shown in Fig.~\ref{LQfigR1}.
To begin, we consider the model of the preceding section involving scalars $R,\widetilde R$ and vectors $V,\widetilde V$, with mixing defined in Eqs.~(\ref{LQmix}-\ref{mixparam}) and couplings to SM fermions given in Eq.~\eqref{LQLint}.  Here, we set to zero $L$-type couplings in Eq.~\eqref{LQLint} and keep only $R$-type ones.  
For each case, the amplitude $a_{RL}^V$ is 
\begin{subequations}
\begin{align}
\textrm{$\mathcal R_{1,2}$-exchange:} \quad a_{RL}^V 
&= - \frac{ \tilde h_R h^*_R \sin 2\theta_R e^{-i\phi_R}}{8\sqrt{2} G_F} \left( \frac{1}{m_{\mathcal{R}_1}^2}
-\frac{1}{m_{\mathcal{R}_2}^2} \right) \\
\textrm{$\mathcal V_{1,2}$-exchange:} \quad a_{RL}^V 
&= \frac{ \tilde g_R g^*_R \sin 2\theta_V e^{-i\phi_V}}{4\sqrt{2} G_F} \left( \frac{1}{m_{\mathcal{V}_1}^2}
-\frac{1}{m_{\mathcal{V}_2}^2} \right) \; .
\end{align}
\end{subequations}
In order to generate $a_{RR}^V$, we introduce two additional LQ states $S$ and $U$, with quantum numbers
\be
\textrm{scalar LQ:} \quad S \sim (\bar{3}, 1, 1/3) \; , \qquad 
\textrm{vector LQ:} \quad U \sim ({3}, 1, 2/3) \; 
\ee
and quark-lepton couplings
\be
\mathscr{L}_{\textrm{int}} = \left( g_S\, \bar{u}^c_R e_R + g_S^\prime \, \bar{d}^c_R \nu_{eR} \right) S 
+ \left( h_U \, \bar{d}_R \gamma^\mu e_R + h_U^\prime \, \bar{u}_R \gamma^\mu \nu_{eR} \right) U_\mu + \; \textrm{h.c.}
\ee
Through tree-level exchange, these states generate
\be
\textrm{$S$-exchange:} \quad a_{RR}^V 
= \frac{g_U^\prime g_U^*}{4\sqrt{2} G_F m_{S}^2}  \, , \qquad 
\textrm{$U$-exchange:} \quad a_{RR}^V 
= - \frac{h_U^\prime h_U^*}{2\sqrt{2} G_F m_{U}^2}  \; .
\ee
There are four possible contributions to  $D_t = \kappa \,\textrm{Im}(a_{RL}^V a_{RR}^{V*})$ depending on which of the combinations
\be \label{3case}
( \mathcal{R}_1, \mathcal{R}_2, S) \, , \quad
( \mathcal{R}_1, \mathcal{R}_2, U)\, , \quad
( \mathcal{V}_1, \mathcal{V}_2, S)\, , \quad
( \mathcal{V}_1, \mathcal{V}_2, U) \; .
\ee
we consider contributing to $a_{RL}^V$ and $a_{RR}^V$.

%---------------------------------------------------------
\begin{figure}[t]
\begin{center}
        \includegraphics[scale=1]{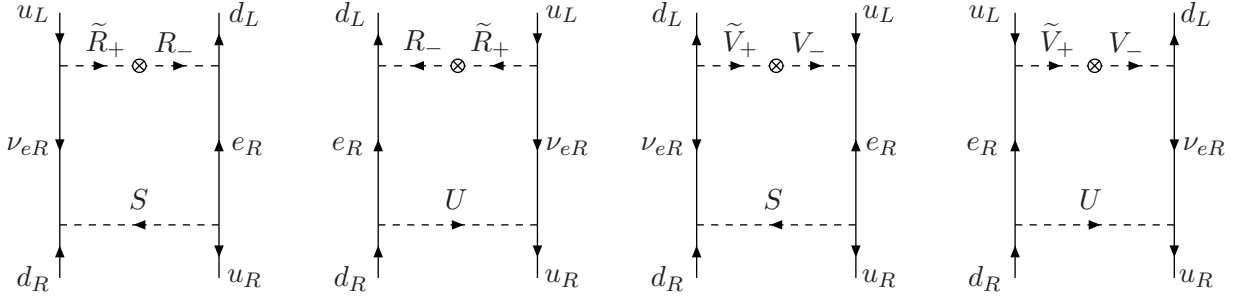} 
\caption{\it LQ contributions to $D_t$ generate radiatively CP-odd operator $\mathcal{O}_{LR}$ contributing to EDMs.}
\label{LQfigR2}
\end{center}
\end{figure}
%%%%%%%%%%%%%%%%%%%%%%%%%%%%%%%%%%%%%%%%%%%%%%5

Next, we consider each of these combinations separately and compute the resulting EDM induced by the CP-odd four quark operator in Eq.~\eqref{Lqq}.  There are four possible contributions, shown in Fig.~\ref{LQfigR2}, and they all give nearly identical results:
\be \label{RHedm}
|k_{LR}| =  \frac{\sqrt{2} G_F m_{LQ}^2}{(4\pi)^2} \,\left|\textrm{Im}(a_{RL}^V a_{RR}^{V*} )\right| \, \hat{f}(m_1^2, m_2^2, m_3^2)
\ee
The loop function is
\be
\hat{f}(m_1^2, m_2^2, m_3^2) \equiv \frac{2\, m_1^2 m_2^2 m_3^2 (m_1^2 \log(m_2^2/m_3^2) + m_2^2 \log(m_3^2/m_1^2) + m_3^2 \log(m_1^2/m_2^2))}{m_{LQ}^2(m_2^2 - m_1^2)(m_3^2-m_2^2)(m_1^2-m_3^2)} 
\ee
where, for each case in Eq.~\eqref{3case}, $m_{1,2,3}^2$ corresponds to the masses of the three states, with $m_{LQ}^2$ being the smallest of the three.  Defined in the way, we have $\hat{f} \ge 1$, with equality if all states are degenerate.  

Assuming that one CP-violating phase is dominant in $D_t$, the resulting EDMs arising from that phase are 
\begin{subequations} \label{LQRbound}
\begin{align}
|d_n| & > 9 \times 10^{-22} \; e \, \textrm{cm} \times \left|{ D_t}/{\kappa} \right| \; \left( \frac{m_{LQ}}{300 \; \textrm{GeV}} \right)^2  \\
|d_{\textrm{Hg}}| & > 7 \times 10^{-26}  \; e \, \textrm{cm}  \times \left|{ D_t}/{\kappa} \right| \; \left( \frac{m_{LQ}}{300 \; \textrm{GeV}} \right)^2  \\
|d_D| & > 4 \times 10^{-22}  \; e \, \textrm{cm}  \times \left|{ D_t}/{\kappa} \right| \; \left( \frac{m_{LQ}}{300 \; \textrm{GeV}} \right)^2 \; .
\end{align}
\end{subequations}
Comparing Eqs. \eqref{DkLR} and \eqref{RHedm}, we find that $D_t$ from LQ scenarios involving right-handed neutrinos is less constrained by EDMs by a factor 4 compared those involving left-handed neutrinos (for fixed $m_{LQ}$).

Constraints on scalar and vector LQ masses from pair production at hadron colliders are the same as in Eqs.~\eqref{cols} and \eqref{colv}.  However, in the limit $h_U \ll h^\prime_U$, the vector $U$ decays primarily via $U \to j \nu$ and is subject to the relatively weaker mass bound $m_U > 144$ GeV~\cite{Abazov:2001mx}.  Significantly stronger bounds are provided by the H1 collaboration for $\beta_e(U) \approx 0$~\cite{Collaboration:2011qa}, which depend on the $U$-$e$-$d$ coupling $h_U$: 
\be
m_U \:\gtrsim \: \left\{ \ba{ccl} 250 \; \textrm{GeV} & \quad & (h_U = 0.03) \\
300 \; \textrm{GeV} & \quad & (h_U = 0.06) \\
1 \; \textrm{TeV} & \quad & (h_U = 0.3) \ea \right.
\ee
Although suppressing $h_U$ weakens the bound on $m_U$, the contribution to $D_t \, (\propto h_U/m_U^2)$ is also suppressed.  Assuming $h_U \gtrsim \mathcal{O}(0.06)$ (to avoid too much additional suppression in $D_t$) we take $m_U \gtrsim 300$ GeV.\footnote{It seems plausible that the best trade-off between small $m_U$ and small $h_U$ occurs for $m_U \sim 300$ GeV, corresponding to the center-of-mass energy $\sqrt{s} = 319$ GeV at HERA.  For $m_U < \sqrt{s}$, on-shell LQ production dominates, allowing for relatively stronger constraints on $h_U$; for $m_U > \sqrt{s}$, only off-shell production is allowed, and the constraints are weaker~\cite{Collaboration:2011qa}.  A more precise analysis is beyond the scope of this work.}
For $m_{LQ} > 300$ GeV, the neutron EDM bound implies $|D_t/\kappa| \lesssim 3 \times 10^{-5}$.

\section{Conclusions}
\label{sec:conclude}

The emiT collaboration has measured $D_n = (-1.0 \pm 2.1)\times 10^{-4}$~\cite{Mumm:2011nd}, consistent the SM prediction dominated by $\mathcal{O}(10^{-5})$ final state effects.  Here, we studied several new physics scenarios beyond the SM and showed that the current neutron EDM measurement $|d_n| < 2.9 \times 10^{-26} \, e \, \textrm{cm}$ provided in all cases stronger bounds on $D$.    
\begin{itemize}
\item $|D_t/\kappa| < 3 \times 10^{-7}$ in left-right symmetric models, exotic fermion models, and the R-parity violating MSSM.  EDM bounds on this class of models, given in Eq.~\eqref{dim6bounds}, can be understood in an otherwise model-independent operator framework through a coupling of the $W$ boson to the right-handed quark charge current $\bar{u}_R \gamma^\mu d_R$.
\item $|D_t/\kappa| < 3 \times 10^{-5}$ ($7 \times 10^{-6}$) in leptoquark models with (without) light right-handed neutrinos.  Moreover, EDM constraints will become more severe if collider bounds on leptoquark masses are improved, as shown in Eqs.~\eqref{LQLbound} and \eqref{LQRbound}.
\end{itemize}
We recall that $\kappa \approx 0.87$ (for the neutron) is defined in Eq.~\eqref{kappa}, and $D_t$ denotes the contribution to $D$ from fundamental T violation (as opposed to final state effects).  Analogous constraints from the mercury EDM bound are weaker by an order of magnitude (with large uncertainties), although the situation may change with future improvements in the nuclear computations.
A future constraint on the deuteron EDM of $|d_D| \lesssim 10^{-28} \, e\, \textrm{cm}$ would improve all aforementioned bounds on $D_t$ by two orders of magnitude.  These bounds can in principle be evaded by fine-tuned cancellations with other CP-odd phases contributing to EDMs, but not to $D_t$.

Even though $D$ is not as sensitive as EDMs to CP violation beyond the SM, clearly it worthwhile to push $D$ measurements to greater sensitivities.  
Since any single EDM measurement has little model discriminating power, it is desirable to consider as many observables as possible --- especially if a non-zero EDM were measured.  
In this case, $D$ could play an important role in untangling the nature of CP violation and potentially shedding light on origin of matter in the Universe.

\begin{acknowledgments}

We thank J.~Behr for many valuable discussions and for encouraging us to pursue this line of research.
We are also indebted to V.~Cirigliano, T.~Chupp, J.~Engel, A.~Garcia, R.~McPherson, M.~Ramsey-Musolf, and I.~Trigger for helpful discussions.  We acknowledge partial support by NSERC of Canada (JN and, previously, ST) and DOE Grant \#DE-FG02-95ER40899 (ST).

\end{acknowledgments}


\begin{thebibliography}{99}

%\cite{Grossman:1997pa}
\bibitem{Grossman:1997pa}
  Y.~Grossman, Y.~Nir, R.~Rattazzi,
  %``CP violation beyond the standard model,''
  Adv.\ Ser.\ Direct.\ High Energy Phys.\  {\bf 15}, 755-794 (1998).
  [hep-ph/9701231].

%\cite{Riotto:1999yt}
\bibitem{Riotto:1999yt}
  A.~Riotto, M.~Trodden,
  %``Recent progress in baryogenesis,''
  Ann.\ Rev.\ Nucl.\ Part.\ Sci.\  {\bf 49}, 35-75 (1999).
  [hep-ph/9901362].

%\cite{Kobayashi:1973fv}
\bibitem{Kobayashi:1973fv}
  M.~Kobayashi, T.~Maskawa,
  %``CP Violation in the Renormalizable Theory of Weak Interaction,''
  Prog.\ Theor.\ Phys.\  {\bf 49}, 652-657 (1973).

%\cite{Jackson:1957zz}
\bibitem{Jackson:1957zz}
  J.~D.~Jackson, S.~B.~Treiman, H.~W.~Wyld,
  %``Possible tests of time reversal invariance in Beta decay,''
  Phys.\ Rev.\  {\bf 106}, 517-521 (1957).

%\cite{Deutsch:1996qm}
\bibitem{Deutsch:1996qm}
  J.~Deutsch, P.~Quin,
  %``Symmetry tests in semileptonic weak interactions: A Search for new physics,''
  In *Langacker, P. (ed.): Precision tests of the standard electroweak model* 706-765.

%\cite{Herczeg:2001vk}
\bibitem{Herczeg:2001vk}
  P.~Herczeg,
  %``Beta decay beyond the standard model,''
  Prog.\ Part.\ Nucl.\ Phys.\  {\bf 46}, 413 (2001).
  %%CITATION = PPNPD,46,413;%%

%\cite{Severijns:2006dr}
\bibitem{Severijns:2006dr}
  N.~Severijns, M.~Beck and O.~Naviliat-Cuncic,
  %``Tests of the standard electroweak model in beta decay,''
  Rev.\ Mod.\ Phys.\  {\bf 78}, 991 (2006)
  [arXiv:nucl-ex/0605029].
  %%CITATION = RMPHA,78,991;%%

%\cite{Nico:2006yg}
\bibitem{Nico:2006yg}
  J.~S.~Nico and W.~M.~Snow,
  %``Experiments in Fundamental Neutron Physics,''
  Ann.\ Rev.\ Nucl.\ Part.\ Sci.\  {\bf 55}, 27 (2005)
  [arXiv:nucl-ex/0612022].
  %%CITATION = ARNUA,55,27;%%

%\cite{Herczeg:1997se}
\bibitem{Herczeg:1997se}
  P.~Herczeg, I.~B.~Khriplovich,
  %``Time reversal violation in Beta decay in the standard model,''
  Phys.\ Rev.\  {\bf D56}, 80-89 (1997).

%\cite{Steinberg:1974zz}
\bibitem{Steinberg:1974zz}
  R.~I.~Steinberg, P.~Liaud, B.~Vignon and V.~W.~Hughes,
  %``New Experimental Limit On T Invariance In Polarized-Neutron Beta Decay,''
  Phys.\ Rev.\ Lett.\  {\bf 33}, 41 (1974).
  %%CITATION = PRLTA,33,41;%%

%\cite{Erozolimsky:1974zz}
\bibitem{Erozolimsky:1974zz}
  B.~G.~Erozolimsky, Y.~A.~Mostovoy, V.~P.~Fedunin {\it et al.},
  %``Continuation of the Search for T Invariance Violation in beta Decay of a Free Neutron,''
  JETP Lett.\  {\bf 20}, 345 (1974).

%\cite{Erozolimsky:1978zi}
\bibitem{Erozolimsky:1978zi}
  B.~G.~Erozolimsky, Y.~.A.~Mostovoi, V.~P.~Fedunin {\it et al.},
  %``False Effects in Investigation of Triple Correlation in Decay of Polarized Neutrons and New Results in Search for Time Parity Violation,''
  Yad.\ Fiz.\  {\bf 28}, 98-104 (1978)

%\cite{Lising:2000pa}
\bibitem{Lising:2000pa}
  L.~J.~Lising {\it et al.}  [emiT Collaboration],
  %``New Limit on the D Coefficient in Polarized Neutron Decay,''
  Phys.\ Rev.\  C {\bf 62}, 055501 (2000)
  [arXiv:nucl-ex/0006001].
  %%CITATION = PHRVA,C62,055501;%%

%\cite{Soldner:2004xm}
\bibitem{Soldner:2004xm}
  T.~Soldner, L.~Beck, C.~Plonka, K.~Schreckenbach and O.~Zimmer,
  %``New limit on T violation in neutron decay,''
  Phys.\ Lett.\  B {\bf 581}, 49 (2004).
  %%CITATION = PHLTA,B581,49;%%

%\cite{Mumm:2011nd}
\bibitem{Mumm:2011nd}
  H.~P.~Mumm, T.~E.~Chupp, R.~L.~Cooper, K.~P.~Coulter, S.~J.~Freedman, B.~K.~Fujikawa, A.~Garcia, G.~L.~Jones {\it et al.},
  %``A New Limit on Time-Reversal Violation in Beta Decay,''
    [arXiv:1104.2778 [nucl-ex]].

%\cite{Baltrusaitis:1977em}
\bibitem{Baltrusaitis:1977em}
  R.~M.~Baltrusaitis, F.~P.~Calaprice,
  %``Improved Experimental Test of Time Reversal Symmetry in Ne-19 beta Decay,''
  Phys.\ Rev.\ Lett.\  {\bf 38}, 464-468 (1977).

%\cite{Hallin:1984mr}
\bibitem{Hallin:1984mr}
  A.~L.~Hallin, F.~P.~Calaprice, D.~W.~Macarthur, L.~E.~Piilonen, M.~B.~Schneider, D.~F.~Schreiber,
  %``Test Of Time Reversal Symmetry In The Beta Decay Of Ne-19,''
  Phys.\ Rev.\ Lett.\  {\bf 52}, 337-340 (1984).

%\cite{Callan:1967zz}
\bibitem{Callan:1967zz}
  C.~G.~Callan, S.~B.~Treiman,
  %``Electromagnetic Simulation of T Violation in Beta Decay,''
  Phys.\ Rev.\  {\bf 162}, 1494-1496 (1967).

%\cite{Ando:2009jk}
\bibitem{Ando:2009jk}
  S.~i.~Ando, J.~A.~McGovern and T.~Sato,
  %``The D coefficient in neutron beta decay in effective field theory,''
  Phys.\ Lett.\  B {\bf 677}, 109 (2009)
  [arXiv:0902.1194 [nucl-th]].
  %%CITATION = PHLTA,B677,109;%%

%\cite{Pospelov:2005pr}
\bibitem{Pospelov:2005pr}
  M.~Pospelov, A.~Ritz,
  %``Electric dipole moments as probes of new physics,''
  Annals Phys.\  {\bf 318}, 119-169 (2005).
  [hep-ph/0504231].

%\cite{Dzuba:2010dy}
\bibitem{Dzuba:2010dy}
  V.~A.~Dzuba and V.~V.~Flambaum,
  %``Current trends in searches for new physics using measurements of parity
  %violation and electric dipole moments in atoms and molecules,''
  arXiv:1009.4960 [physics.atom-ph].
  %%CITATION = ARXIV:1009.4960;%%

%\cite{Baker:2006ts}
\bibitem{Baker:2006ts}
  C.~A.~Baker {\it et al.},
  %``An improved experimental limit on the electric dipole moment of the
  %neutron,''
  Phys.\ Rev.\ Lett.\  {\bf 97}, 131801 (2006)
  [arXiv:hep-ex/0602020].
  %%CITATION = PRLTA,97,131801;%%

%\cite{Griffith:2009zz}
\bibitem{Griffith:2009zz}
  W.~C.~Griffith, M.~D.~Swallows, T.~H.~Loftus, M.~V.~Romalis, B.~R.~Heckel and E.~N.~Fortson,
  %``Improved Limit on the Permanent Electric Dipole Moment of Hg-199,''
  Phys.\ Rev.\ Lett.\  {\bf 102}, 101601 (2009).
  %%CITATION = PRLTA,102,101601;%%

%\cite{Regan:2002ta}
\bibitem{Regan:2002ta}
  B.~C.~Regan, E.~D.~Commins, C.~J.~Schmidt and D.~DeMille,
  %``New limit on the electron electric dipole moment,''
  Phys.\ Rev.\ Lett.\  {\bf 88}, 071805 (2002).
  %%CITATION = PRLTA,88,071805;%%

%\cite{Hudson:2011zz}
\bibitem{Hudson:2011zz}
  J.~J.~Hudson, D.~M.~Kara, I.~J.~Smallman, B.~E.~Sauer, M.~R.~Tarbutt, E.~A.~Hinds,
  %``Improved measurement of the shape of the electron,''
  Nature {\bf 473}, 493-496 (2011).

%\cite{Pospelov:1999ha}
\bibitem{Pospelov:1999ha}
  M.~Pospelov and A.~Ritz,
  %``Theta-Induced Electric Dipole Moment of the Neutron via QCD Sum Rules,''
  Phys.\ Rev.\ Lett.\  {\bf 83}, 2526 (1999)
  [arXiv:hep-ph/9904483].
  %%CITATION = PRLTA,83,2526;%%

%\cite{Ellis:2008zy}
\bibitem{Ellis:2008zy}
  J.~R.~Ellis, J.~S.~Lee, A.~Pilaftsis,
  %``Electric Dipole Moments in the MSSM Reloaded,''
  JHEP {\bf 0810}, 049 (2008).
  [arXiv:0808.1819 [hep-ph]].

%\cite{Li:2010ax}
\bibitem{Li:2010ax}
  Y.~Li, S.~Profumo, M.~Ramsey-Musolf,
  %``A Comprehensive Analysis of Electric Dipole Moment Constraints on CP-violating Phases in the MSSM,''
  JHEP {\bf 1008}, 062 (2010).
  [arXiv:1006.1440 [hep-ph]].

%\cite{LRsym}
\bibitem{LRsym}
  J.~C.~Pati, A.~Salam,
  %``Lepton Number as the Fourth Color,''
  Phys.\ Rev.\  {\bf D10}, 275-289 (1974).
  R.~N.~Mohapatra, J.~C.~Pati,
  %``Left-Right Gauge Symmetry and an Isoconjugate Model of CP Violation,''
  Phys.\ Rev.\  {\bf D11}, 566-571 (1975).  {\it ibid.} Phys.\ Rev.\  {\bf D11}, 2558 (1975).
  G.~Senjanovic, R.~N.~Mohapatra,
  %``Exact Left-Right Symmetry and Spontaneous Violation of Parity,''
  Phys.\ Rev.\  {\bf D12}, 1502 (1975).

%\cite{Barbier:2004ez}
\bibitem{Barbier:2004ez}
  R.~Barbier, C.~Berat, M.~Besancon, M.~Chemtob, A.~Deandrea, E.~Dudas, P.~Fayet, S.~Lavignac {\it et al.},
  %``R-parity violating supersymmetry,''
  Phys.\ Rept.\  {\bf 420}, 1-202 (2005).
  [hep-ph/0406039].

%\cite{Langacker:1988ur}
\bibitem{Langacker:1988ur}
  P.~Langacker, D.~London,
  %``Mixing Between Ordinary and Exotic Fermions,''
  Phys.\ Rev.\  {\bf D38}, 886 (1988).

%\cite{Buchmuller:1986zs}
\bibitem{Buchmuller:1986zs}
  W.~Buchmuller, R.~Ruckl, D.~Wyler,
  %``Leptoquarks in Lepton - Quark Collisions,''
  Phys.\ Lett.\  {\bf B191}, 442-448 (1987).

%\cite{Herczeg:1982ij}
\bibitem{Herczeg:1982ij}
  P.~Herczeg,
  %``On The Mohapatra-pati Model Of Cp Violation,''
  Phys.\ Rev.\  {\bf D28}, 200 (1983).

%\cite{Herczeg:2005yf}
\bibitem{Herczeg:2005yf}
  P.~Herczeg,
  %``The T-odd R and D correlations in beta decay,''
  J.\ Res.\ Natl.\ Inst.\ Stand.\ Tech.\  {\bf 110}, 453-459 (2005).

%\cite{An:2009zh}
\bibitem{An:2009zh}
  H.~An, X.~Ji, F.~Xu,
  %``P-odd and CP-odd Four-Quark Contributions to Neutron EDM,''
  JHEP {\bf 1002}, 043 (2010).
  [arXiv:0908.2420 [hep-ph]].

%\cite{Ban:2010ea}
\bibitem{Ban:2010ea}
  S.~Ban, J.~Dobaczewski, J.~Engel and A.~Shukla,
  %``Fully self-consistent calculations of nuclear Schiff moments,''
  Phys.\ Rev.\  C {\bf 82}, 015501 (2010)
  [arXiv:1003.2598 [nucl-th]].
  %%CITATION = PHRVA,C82,015501;%%


%\cite{Kurylov:2001zx}
\bibitem{Kurylov:2001zx}
  A.~Kurylov, M.~J.~Ramsey-Musolf,
  %``Charged current universality in the MSSM,''
  Phys.\ Rev.\ Lett.\  {\bf 88}, 071804 (2002).
  [hep-ph/0109222].
  J.~Erler, M.~J.~Ramsey-Musolf,
  %``Low energy tests of the weak interaction,''
  Prog.\ Part.\ Nucl.\ Phys.\  {\bf 54}, 351-442 (2005).
  [hep-ph/0404291].
  A.~Czarnecki, W.~J.~Marciano, A.~Sirlin,
  %``Precision measurements and CKM unitarity,''
  Phys.\ Rev.\  {\bf D70}, 093006 (2004).
  [hep-ph/0406324].
  I.~S.~Towner, J.~C.~Hardy,
  %``The evaluation of V(ud) and its impact on the unitarity of the Cabibbo-Kobayashi-Maskawa quark-mixing matrix,''
  Rept.\ Prog.\ Phys.\  {\bf 73}, 046301 (2010).


%\cite{Kopp:2011qd}
\bibitem{Kopp:2011qd}
  J.~Kopp, M.~Maltoni, T.~Schwetz,
  %``Are there sterile neutrinos at the eV scale?,''
  Phys.\ Rev.\ Lett.\  {\bf 107}, 091801 (2011).
  [arXiv:1103.4570 [hep-ph]].
  C.~Giunti, M.~Laveder,
  %``Towards 3+1 Neutrino Mixing,''
  [arXiv:1109.4033 [hep-ph]].

%\cite{Hamann:2011ge}
\bibitem{Hamann:2011ge}
  J.~Hamann, S.~Hannestad, G.~G.~Raffelt, Y.~Y.~Y.~Wong,
  %``Sterile neutrinos with eV masses in cosmology: How disfavoured exactly?,''
  JCAP {\bf 1109}, 034 (2011).
  [arXiv:1108.4136 [astro-ph.CO]].


%\cite{Nakamura:2010zzi}
\bibitem{Nakamura:2010zzi}
  K.~Nakamura {\it et al.}  [Particle Data Group],
  %``Review of particle physics,''
  J.\ Phys.\ G {\bf 37}, 075021 (2010).
  %%CITATION = JPHGB,G37,075021;%%

%\cite{Adler:1975he}
\bibitem{Adler:1975he}
  S.~L.~Adler, E.~W.~.~Colglazier, J.~B.~Healy, I.~Karliner, J.~Lieberman, Y.~J.~Ng and H.~S.~Tsao,
  %``RENORMALIZATION CONSTANTS FOR SCALAR, PSEUDOSCALAR, AND TENSOR CURRENTS,''
  Phys.\ Rev.\  D {\bf 11}, 3309 (1975).
  %%CITATION = PHRVA,D11,3309;%%

%\cite{Bhattacharya:2011qm}
\bibitem{Bhattacharya:2011qm}
  T.~Bhattacharya {\it et al.},
  %``Probing Novel Scalar and Tensor Interactions from (Ultra)Cold Neutrons to
  %the LHC,''
  arXiv:1110.6448 [hep-ph].
  %%CITATION = ARXIV:1110.6448;%%

%\cite{MP:1991}
\bibitem{MP:1991}
  A.-M.~M\r{a}rtensson-Pendrill and E.~Lindroth, Europhys.\ Lett.\ {\bf 15}, 155 (1991).

%\cite{Khriplovich:1997ga}
\bibitem{Khriplovich:1997ga}
  I.~B.~Khriplovich and S.~K.~Lamoreaux,
  %``CP Violation Without Strangeness: Electric Dipole Moments Of Particles,
  %Atoms, And Molecules,''
%\href{http://www.slac.stanford.edu/spires/find/hep/www?irn=4989619}{SPIRES entry}
{\it  Berlin, Germany: Springer (1997) 230 p}

%\cite{Ginges:2003qt}
\bibitem{Ginges:2003qt}
  J.~S.~M.~Ginges, V.~V.~Flambaum,
  %``Violations of fundamental symmetries in atoms and tests of unification theories of elementary particles,''
  Phys.\ Rept.\  {\bf 397}, 63-154 (2004).
  [physics/0309054].

%\cite{Sahoo:2008}
\bibitem{Sahoo:2008}
  B.~K.~Sahoo, B.~P.~Das, R.~K.~Chaudhuri, D.~Mukherjee, E.~P.~Venugopal,
  Phys.\ Rev.\ A {\bf 78}, 010501(R) (2008).

%\cite{Dzuba:2009mw}
\bibitem{Dzuba:2009mw}
  V.~A.~Dzuba, V.~V.~Flambaum,
  %``Calculation of the (T,P)-odd Electric Dipole Moment of Thallium,''
  Phys.\ Rev.\  {\bf A80}, 062509 (2009).
  [arXiv:0909.0308 [physics.atom-ph]].

%\cite{Weinberg:1989dx}
\bibitem{Weinberg:1989dx}
  S.~Weinberg,
%  ``Larger Higgs Exchange Terms in the Neutron Electric Dipole Moment,''
  Phys.\ Rev.\ Lett.\  {\bf 63}, 2333 (1989).
  %%CITATION = PRLTA,63,2333;%%

%\cite{Khatsimovsky:1987fr}
\bibitem{Khatsimovsky:1987fr}
  V.~M.~Khatsimovsky, I.~B.~Khriplovich and A.~S.~Yelkhovsky,
  %``NEUTRON ELECTRIC DIPOLE MOMENT, T ODD NUCLEAR FORCES AND NATURE OF CP
  %VIOLATION,''
  Annals Phys.\  {\bf 186}, 1 (1988).

%\cite{He:1989xj}
\bibitem{He:1989xj}
  X.~-G.~He, B.~H.~J.~McKellar, S.~Pakvasa,
  %``The Neutron Electric Dipole Moment,''
  Int.\ J.\ Mod.\ Phys.\  {\bf A4}, 5011 (1989).
  %%CITATION = APNYA,186,1;%%

%\cite{He:1992db}
\bibitem{He:1992db}
  X.~G.~He and B.~McKellar,
  %``Large Contribution To The Neutron Electric Dipole Moment From A
  %Dimension-Six Four Quark Operator,''
  Phys.\ Rev.\  D {\bf 47}, 4055 (1993).
  %%CITATION = PHRVA,D47,4055;%%

%\cite{Xu:2009nt}
\bibitem{Xu:2009nt}
  F.~Xu, H.~An, X.~Ji,
  %``Neutron Electric Dipole Moment Constraint on Scale of Minimal Left-Right Symmetric Model,''
  JHEP {\bf 1003}, 088 (2010).
  [arXiv:0910.2265 [hep-ph]].

%\cite{Semertzidis:2009zza}
\bibitem{Semertzidis:2009zza}
  Y.~K.~Semertzidis,
  %``Review Of Electric Dipole Moments Of Fundamental Particles,''
  AIP Conf.\ Proc.\  {\bf 1149}, 48 (2009).
  %%CITATION = APCPC,1149,48;%%

%\cite{Khatsymovsky:1992yg}
\bibitem{Khatsymovsky:1992yg}
  V.~M.~Khatsymovsky, I.~B.~Khriplovich,
  %``CP odd interaction of light quarks and the neutron electric dipole moment,''
  Phys.\ Lett.\  {\bf B296}, 219-222 (1992).

%\cite{Dzuba:2009kn}
\bibitem{Dzuba:2009kn}
  V.~A.~Dzuba, V.~V.~Flambaum, S.~G.~Porsev,
  %``Calculation of P,T-odd electric dipole moments for diamagnetic atoms Xe-129, Yb-171, Hg-199, Rn-211, and Ra-225,''
  Phys.\ Rev.\ A {\bf 80}, 032120 (2009).
  [arXiv:0906.5437 [physics.atom-ph]].

%\cite{Dzuba:2002kg}
\bibitem{Dzuba:2002kg}
  V.~A.~Dzuba, V.~V.~Flambaum, J.~S.~M.~Ginges, M.~G.~Kozlov,
  %``Electric dipole moments of Hg, Xe, Rn, Ra, Pu, and TlF induced by the nuclear Schiff moment and limits on time reversal violating interactions,''
  Phys.\ Rev.\  {\bf A66}, 012111 (2002).
  [hep-ph/0203202].

%\cite{Latha:2009nq}
\bibitem{Latha:2009nq}
  K.~V.~P.~Latha, D.~Angom, B.~P.~Das, D.~Mukherjee,
  %``Probing CP violation with the electric dipole moment of atomic mercury,''
  Phys.\ Rev.\ Lett.\  {\bf 103}, 083001 (2009).
  [arXiv:0902.4790 [physics.atom-ph]].

%\cite{Herczeg:1987}
\bibitem{Herczeg:1987}
  P.~Herczeg,
  In *Chapel Hill 1987, Proceedings, Tests of time reversal invariance in neutron physics* 24-53.



%\cite{deJesus:2005nb}
\bibitem{deJesus:2005nb}
  J.~H.~de Jesus and J.~Engel,
  %``Time-Reversal-Violating Schiff Moment of 199Hg,''
  Phys.\ Rev.\  C {\bf 72}, 045503 (2005)
  [arXiv:nucl-th/0507031].
  %%CITATION = PHRVA,C72,045503;%%

%\cite{Dmitriev:2003kb}
\bibitem{Dmitriev:2003kb}
  V.~F.~Dmitriev and R.~A.~Sen'kov,
  %``P- and T-violating Schiff moment of the Mercury nucleus,''
  Phys.\ Atom.\ Nucl.\  {\bf 66}, 1940 (2003)
  [Yad.\ Fiz.\  {\bf 66}, 1988 (2003)]
  [arXiv:nucl-th/0304048].
  %%CITATION = YAFIA,66,1988;%%


%\cite{Ellis:2011hp}
\bibitem{Ellis:2011hp}
  J.~Ellis, J.~S.~Lee, A.~Pilaftsis,
  %``Maximal Electric Dipole Moments of Nuclei with Enhanced Schiff Moments,''
  JHEP {\bf 1102}, 045 (2011).
  [arXiv:1101.3529 [hep-ph]].

%\cite{He:1992jh}
\bibitem{He:1992jh}
  X.~-G.~He, B.~McKellar,
  %``Constraints on CP violating nucleon-nucleon interactions in gauge models from atomic electric dipole moment,''
  Phys.\ Rev.\  {\bf D46}, 2131-2140 (1992).


%\cite{deVries:2011an}
\bibitem{deVries:2011an}
  J.~de Vries, R.~Higa, C.~-P.~Liu, E.~Mereghetti, I.~Stetcu, R.~G.~E.~Timmermans, U.~van Kolck,
  %``Electric Dipole Moments of Light Nuclei From Chiral Effective Field Theory,''
  [arXiv:1109.3604 [hep-ph]].


%\cite{Khriplovich:1999qr}
\bibitem{Khriplovich:1999qr}
  I.~B.~Khriplovich and R.~A.~Korkin,
  %``P and T odd electromagnetic moments of deuteron in chiral limit,''
  Nucl.\ Phys.\  A {\bf 665}, 365 (2000)
  [arXiv:nucl-th/9904081].
  %%CITATION = NUPHA,A665,365;%%

%\cite{Liu:2004tq}
\bibitem{Liu:2004tq}
  C.~P.~Liu and R.~G.~E.~Timmermans,
  %``P- and T-odd two-nucleon interaction and the deuteron electric dipole
  %moment,''
  Phys.\ Rev.\  C {\bf 70}, 055501 (2004)
  [arXiv:nucl-th/0408060].
  %%CITATION = PHRVA,C70,055501;%%

%\cite{Korkin:2005bw}
\bibitem{Korkin:2005bw}
  R.~V.~Korkin,
  %``P and T odd effects in deuteron in the Reid potential,''
  arXiv:nucl-th/0504078.
  %%CITATION = NUCL-TH/0504078;%%

%\cite{Afnan:2010xd}
\bibitem{Afnan:2010xd}
  I.~R.~Afnan and B.~F.~Gibson,
  %``Model Dependence of the 2H Electric Dipole Moment,''
  Phys.\ Rev.\  C {\bf 82}, 064002 (2010)
  [arXiv:1011.4968 [nucl-th]].
  %%CITATION = PHRVA,C82,064002;%%

%\cite{Buchmuller:1985jz}
\bibitem{Buchmuller:1985jz}
  W.~Buchmuller, D.~Wyler,
  %``Effective Lagrangian Analysis of New Interactions and Flavor Conservation,''
  Nucl.\ Phys.\  {\bf B268}, 621 (1986).

%\cite{Drees:2003dv}
\bibitem{Drees:2003dv}
  M.~Drees, M.~Rauch,
  %``Complete one loop calculation of the T violating D parameter in neutron decay in the MSSM,''
  Eur.\ Phys.\ J.\  {\bf C29}, 573-585 (2003).
  [hep-ph/0302244].


%\cite{Yamanaka:2009hi}
\bibitem{Yamanaka:2009hi}
  N.~Yamanaka, T.~Sato, T.~Kubota,
  %``R-parity violating supersymmetric contributions to the neutron beta decay,''
  J.\ Phys.\ G {\bf G37}, 055104 (2010).
  [arXiv:0908.1007 [hep-ph]].

%\cite{Martin:1997ns}
\bibitem{Martin:1997ns}
  S.~P.~Martin,
  %``A Supersymmetry primer,''
  In *Kane, G.L. (ed.): Perspectives on supersymmetry* 1-98.
  [arXiv:hep-ph/9709356 [hep-ph]].


%\cite{Chang:1996sw}
\bibitem{Chang:1996sw}
  D.~Chang, W.~-Y.~Keung,
  %``New limits on R-parity breakings in supersymmetric standard models,''
  Phys.\ Lett.\  {\bf B389}, 294-298 (1996).
  [hep-ph/9608313].


%\cite{Godbole:1999ye}
\bibitem{Godbole:1999ye}
  R.~M.~Godbole, S.~Pakvasa, S.~D.~Rindani, X.~Tata,
  %``Fermion dipole moments in supersymmetric models with explicitly broken R-parity,''
  Phys.\ Rev.\  {\bf D61}, 113003 (2000).
  [hep-ph/9912315].


%\cite{Georgi:1974sy}
\bibitem{Georgi:1974sy}
  H.~Georgi, S.~L.~Glashow,
  %``Unity of All Elementary Particle Forces,''
  Phys.\ Rev.\ Lett.\  {\bf 32}, 438-441 (1974).
  J.~C.~Pati, A.~Salam,
  %``Lepton Number as the Fourth Color,''
  Phys.\ Rev.\  {\bf D10}, 275-289 (1974).

%\cite{Schrempp:1984nj}
\bibitem{Schrempp:1984nj}
  B.~Schrempp, F.~Schrempp,
  %``Light Leptoquarks,''
  Phys.\ Lett.\  {\bf B153}, 101 (1985).
  J.~Wudka,
  %``Composite Leptoquarks,''
  Phys.\ Lett.\  {\bf B167}, 337 (1986).

%\cite{Khriplovich:1990ef}
\bibitem{Khriplovich:1990ef}
  I.~B.~Khriplovich,
  %``What do we know about T odd but P even interaction?,''
  Nucl.\ Phys.\  {\bf B352}, 385-401 (1991).

%\cite{Kurylov:2000ub}
\bibitem{Kurylov:2000ub}
  A.~Kurylov, G.~C.~McLaughlin, M.~J.~Ramsey-Musolf,
  %``Constraints on T odd, P even interactions from electric dipole moments, revisited,''
  Phys.\ Rev.\  {\bf D63}, 076007 (2001).
  [hep-ph/0011185].

%\cite{Khachatryan:2010mp}
\bibitem{Khachatryan:2010mp}
  V.~Khachatryan {\it et al.} [ CMS Collaboration ],
  %``Search for Pair Production of First-Generation Scalar Leptoquarks in pp Collisions at sqrt(s) = 7 TeV,''
  Phys.\ Rev.\ Lett.\  {\bf 106}, 201802 (2011).
  [arXiv:1012.4031 [hep-ex]]
  S.~Chatrchyan {\it et al.} [ CMS Collaboration ],
  %``Search for First Generation Scalar Leptoquarks in the evjj channel in pp collisions at sqrt(s) = 7 TeV,''  
  [arXiv:1105.5237 [hep-ex]].


%\cite{Aad:2011uv}
\bibitem{Aad:2011uv}
  G.~Aad {\it et al.} [ ATLAS Collaboration ],
  %``Search for pair production of first or second generation leptoquarks in proton-proton collisions at sqrt(s)=7 TeV using the ATLAS detector at the LHC,''
  [arXiv:1104.4481 [hep-ex]]

%\cite{Abazov:2011qj}
\bibitem{Abazov:2011qj}
  V.~M.~Abazov {\it et al.} [ D0 Collaboration ],
  %``Search for first generation leptoquark pair production in the electron + missing energy + jets final state,''  
  [arXiv:1107.1849 [hep-ex]].



%\cite{D0:2009gf}
\bibitem{D0:2009gf}
  V.~M.~Abazov {\it et al.} [ D0 Collaboration ],
  %``Search for pair production of first-generation leptoquarks in p anti-p collisions at s**(1/2) = 1.96-TeV,''
  Phys.\ Lett.\  {\bf B681}, 224-232 (2009).
  [arXiv:0907.1048 [hep-ex]].

%\cite{Aad:2011ib}
\bibitem{Aad:2011ib}
  G.~Aad {\it et al.} [ ATLAS Collaboration ],
  %``Search for squarks and gluinos using final states with jets and missing transverse momentum with the ATLAS detector in sqrt(s) = 7 TeV proton-proton collisions,''
    [arXiv:1109.6572 [hep-ex]].



%\cite{Abazov:2001mx}
\bibitem{Abazov:2001mx}
  V.~M.~Abazov {\it et al.} [ D0 Collaboration ],
  %``Search for first generation scalar and vector leptoquarks,''
  Phys.\ Rev.\  {\bf D64}, 092004 (2001).
  [hep-ex/0105072].


%\cite{Collaboration:2011qa}
\bibitem{Collaboration:2011qa}
  H1~Collaboration,
  %``Search for First Generation Leptoquarks in ep Collisions at HERA,''
  [arXiv:1107.3716 [hep-ex]].



\end{thebibliography}
\end{document}